\RequirePackage{fixltx2e}
\documentclass[aip,jcp,preprint,floatfix]{revtex4-1}

\usepackage{graphicx} 
\usepackage{color}
\usepackage{pagecolor,lipsum}% http://ctan.org/pkg/{pagecolor,lipsum}
\usepackage{dcolumn}
\usepackage{amsmath,amsfonts} %
\usepackage[acronym]{glossaries} %
\usepackage[breaklinks, % 
colorlinks, % 
linkcolor=blue, % 
citecolor=blue, % 
urlcolor=blue]{hyperref}

\newacronym{CI}{CI}{conical intersection}
\newacronym{GP}{GP}{geometric phase}
\newacronym{DBOC}{DBOC}{diagonal Born-Oppenheimer correction}
\newacronym{MCA}{MCA}{moving crude adiabatic}
\newacronym{BMA}{BMA}{bis(methylene) adamantyl cation}
\newacronym{MCTDH}{MCTDH}{multi configuration time dependent Hartree}
\newcommand{\ket}[1]{\ensuremath{\left\vert #1\right\rangle}}

\newcommand{\braket}[2]{\ensuremath{\left\langle #1\vert#2\right\rangle}}
\newcommand{\braOket}[3]{\ensuremath{\left\langle #1 \left\vert #2 \right\vert #3 \right\rangle}}
\newcommand{\imag}{i}
\begin{document}

%%%% Article title to be placed here
\title{The moving crude adiabatic alternative to the adiabatic representation in excited state dynamics.}

% Employing adiabatic states in non-adiabatic dynamics with a Gaussian basis: Using moving crude adiabatic states to describe non-adiabatic transitions.
% Can we utilize adiabatic states in non-adiabatic on-the-fly dynamics?
% The moving crude adiabatic alternative to the adiabatic representation in excited state dynamics.
% The moving crude adiabatic to reconcile Gaussian-based direct non-adiabatic methods with adiabatic states.
% 

\author{R. Maskri}
\affiliation{MSME, Univ Gustave Eiffel, CNRS UMR 8208, Univ Paris Est Creteil, F-77474 Marne-la-Vall\'ee, France}

\author{L. Joubert-Doriol}
\email{loic.joubert-doriol@univ-eiffel.fr}
\affiliation{MSME, Univ Gustave Eiffel, CNRS UMR 8208, Univ Paris Est Creteil, F-77474 Marne-la-Vall\'ee, France}

%%%% Abstract text to be placed here %%%%%%%%%%%%
\begin{abstract}
The choice of the electronic representation in on-the-fly quantum dynamics is crucial.
The adiabatic representation is appealing since adiabatic states are readily available from quantum chemistry packages.
The nuclear wavepackets are then expanded in a basis of Gaussian functions, which follow trajectories to explore the potential energy surfaces and approximate the potential using a local expansion of the adiabatic quantities.
Nevertheless, the adiabatic representation is plagued with severe limitations when conical intersections are involved: the \glspl{DBOC} are non-integrable, and the geometric phase effect on the nuclear wavepackets cannot be accounted for unless a model is available.
To circumvent these difficulties, the \gls{MCA} representation was proposed and successfully tested in low energy dynamics where the wavepacket skirts the conical intersection.
We assess the \gls{MCA} representation in the case of non-adiabatic transitions through conical intersections.
First, we show that using a Gaussian basis in the adiabatic representation indeed exhibits the aforementioned difficulties with a special emphasis on the possibility to regularize the \gls{DBOC} terms.
Then, we show that \gls{MCA} is indeed able to properly model non-adiabatic transitions.
Tests are done on linear vibronic coupling models for the bis(methylene) adamantyl cation and the butatriene cation.
\end{abstract}
%%%%%%%%%%%%%%%%%%%%%%%%%%%

\date{\today}

\maketitle

\glsresetall

%%%%%%%%%%%%%%%%%%%%%%%%%%%%%%%%%%%%%%%%%%%%%%%%%%%%%%%%%%%%%%%%%%%%%%%%%%%%%%%%%%
\section{Introduction}
%%%% Insert A head here

The Born-Oppenheimer approximation is a cornerstone of chemistry that greatly simplifies the analysis of molecular systems by defining electronic adiabatic states and confining the system in one of these states.
Nevertheless, this approximation regularly fails in describing photochemistry, particularly in regions of the potential energy surfaces where the adiabatic states become degenerate and form the commonly encountered \glspl{CI}~\cite{Yarkoni:2001/jpca/6277,Migani:2004/Book}.
Indeed, the non-adiabatic couplings that are neglected in the Born-Oppenheimer approximation can become arbitrarily large close to a \gls{CI}~\cite{Migani:2004/Book,Baer:2006/Book,Yarkoni:2004/Book} and break the Born-Oppenheimer approximation.
In this case one may believe that including the missing couplings should be sufficient to recover the exact description, but this is not completely true.
A first difficulty comes from the fact that the non-adiabatic couplings diverge at the \glspl{CI}.
Hence, all the non-adiabatic coupling terms may not be integrable over the chosen basis and can lead to numerical instabilities.
Another difficulty is that an extra \gls{GP} attached to the adiabatic states appears upon encircling the \glspl{CI}~\cite{Berry:1984/prsa/45,Mead:1979/jcp/2284,Mead:1980/cp/23,Berry:1987/prsa/31}.
The \gls{GP} implies that adiabatic states have a double-valued boundary condition when revolving around the \gls{CI}, and so must be the boundary condition on the nuclear wavepackets in order to recover a single valued total molecular wavefunction~\cite{Mead:1979/jcp/2284,Longuet:1958/prsa/1}.
Overlooking this \gls{GP} was found to severely impact the resulting dynamics~\cite{Ryabinkin:2017/acr/1785,Xie:2016/jacs/7828,Henshaw:2017/jpcl/146,Neville:2020/arxiv,Xie:2019/acr/501}.
Not accounting for \gls{GP} diminishes the transition probability in internal conversion processes~\cite{Ryabinkin:2014/jcp/214116,Li:2017/jcp/064106}.
Even in a low energy regime, when the nuclear wavepacket encircles but does not necessarily have enough energy to reach the \gls{CI}, the extra \gls{GP}-induced interference causes appearance of a nodal plane in the nuclear wavepacket~\cite{Schon:1995/jcp/9292,Ryabinkin:2013/prl/220406,Xie:2016/jacs/7828} that can diminish population transfer around a \gls{CI} or even localize the wavepacket on one side of the \gls{CI}~\cite{Ryabinkin:2013/prl/220406,Joubert:2013/jcp/234103,Joubert:2017/cc/7365,Henshaw:2017/jpcl/146}.

Non-adiabatic dynamics is often described using quantum-classical methods, such as surface hopping~\cite{tully:1998/fd/407} or using the exact factorization approach~\cite{Min:2015/prl/073001}.
Regions containing \glspl{CI} are known to exhibit strong quantum effects due to the strong electron-nuclear couplings, for instance the aforementioned \gls{GP}-induced interference~\cite{Ryabinkin:2013/prl/220406,Joubert:2013/jcp/234103,Xie:2016/jacs/7828}.
A full quantum approach is then necessary to properly account for these quantum effects.
Usual approaches for such simulations use a multiconfiguration expansion in combination with multidimensional grids (see for example the \gls{MCTDH} method~\cite{Meyer:1990/cpl/73,Meyer:2009/Book}).
Some drawback of these methods are the necessity of generating a model Hamiltonian beforehand, and the fact that grid-based methods scale exponentially with the number of dimensions.
Both limitations can be addressed using another common representation of the nuclear wavepackets employing an expansion in terms of time-dependent Gaussian functions~\cite{Heller:1975/jcp/1544,Heller:1981/jcp/2923,Burghardt:1999/jcp/2927,Ben:2000/jpca/5161,Worth:2004/fd/307,Saita:2012/jcp/22A506,Neville:2016/fd/117,Meek:2016/jcp/184103,Joubert:2018/jpca/6031}.
This representation has the strong advantage that it avoids evaluation of the Hamiltonian integrals over multidimensional grid by using Gaussian integration when possible, or by utilizing a local approximation motivated by the fact that Gaussian functions have a limited spatial extension.
This approach is particularly useful for on-the-fly simulation of molecular systems for which the potential energy surfaces are not known analytically (see for instance the integrals' approximations described in Refs.~\citenum{Mignolet:2018/jcp/134110,Makhov:2017/cp/200,Richings:2015/irpc/269}).
However, a severe drawback of the Gaussian expansion is that Gaussian functions are not double-valued functions.
A general approach to solve this difficulty is to remove the \gls{GP} from the adiabatic states by using an extra phase factor~\cite{Mead:1979/jcp/2284}.
Nevertheless, choosing the right phase factor requires prior knowledge of the \glspl{CI} locations.
However, their locations are generally not known in on-the-fly simulations of real molecular systems.
Thus, it is not always possible to cancel the \gls{GP} of the adiabatic states~\cite{Xie:2017/jcp/044109}.
Another drawback comes from the fact that the diagonal elements of the non-adiabatic coupling matrix, also known as \glspl{DBOC}, are not often calculated in electronic structure packages and are furthermore non-integrable unless special basis functions are utilized~\cite{Meek:2016/jcp/184109,Fedorov:2019/jcp/054102}.
Thus, \glspl{DBOC} are also often overlooked.
Despite these difficulties, time-dependent Gaussian functions are very often employed in non-adiabatic dynamics for the advantages they provide in on-the-fly quantum dynamics.

The difficulties appearing at \glspl{CI} can in principle be resolved by applying a diabatization of the electronic states~\cite{Koppel:2004/Book}.
However, only approximate quasi-diabatization are possible~\cite{Meek:2016/jcp/184103,Richings:2015/irpc/269,Koppel:2006/mp/1069,Richings:2017/cpl/606}, which introduce an additional source of error.
Here, we focus on an alternative representation of the electronic states: the \gls{MCA} representation~\cite{Fernandez:2016/pccp/10028,Joubert:2017/jpcl/452,Joubert:2018/jcp/114102}.
In the \gls{MCA} representation, the electronic states are solution of the electronic time-independent Schr\"odinger equation only at the Gaussians' center.
Such electronic states do not depend on the nuclear coordinates anymore but rather on the Gaussians' center that are simply time-dependent parameters.
Consequently, these states are time-dependent diabatic states, and they do not exhibit \gls{GP} by construction (they are single-valued, as opposed to the adiabatic states).
Hence, the nuclear components are also single-valued, and the single-valuedness of the Gaussian basis functions does not pose any problem (in opposition to the adiabatic case).
Using \gls{MCA} states was found to properly account for \gls{GP} effects in low energy dynamics in adiabatic representation~\cite{Joubert:2017/jpcl/452}, where the extra \gls{GP}-induced interference causes appearance of a nodal plane in the adiabatic nuclear wavepackets~\cite{Ryabinkin:2013/prl/220406,Joubert:2017/jpcl/452}.
On contrary, the \gls{MCA} states were not yet tested in the regime of non-adiabatic transitions, where \gls{GP} participates in enhancing interstate population transfer in the adiabatic representation.
Furthermore, the \gls{MCA} states do not produce non-adiabatic couplings, including \gls{DBOC}, which is also a powerful advantage for the numerical stability of the simulations.

In this paper, we study the ability of the \gls{MCA} representation to properly describe the non-adiabatic transitions with a special emphasis on the role of \gls{GP} and \gls{DBOC} in the adiabatic representation.
The role of \gls{GP} in this context was found to be twofold: i) it `compensates for repulsion caused by the \gls{DBOC}'~\cite{Ryabinkin:2014/jcp/214116} and thus facilitates the wavefunction to reach the \gls{CI} region of large non-adiabatic couplings, and ii) `it enhances transfer probability for a component of a nuclear wave-packet that corresponds to the zero eigenvalue of the [angular momentum] operator defined with respect to the \gls{CI} point'~\cite{Ryabinkin:2014/jcp/214116}.
The first effect, compensation of \gls{DBOC}, was illustrated on the electron transfer dynamics of the \gls{BMA}, while the second, enhancement of probability transfer, was best illustrated by the dynamics of the butatriene cation.
In the spirit of ref.~\citenum{Ryabinkin:2014/jcp/214116}, we start by exhibiting that employing double-valued adiabatic states with nuclear components that are single-valued overlooks the \gls{GP} such that the resulting dynamics can deviate from the exact result.
Then, we numerically expose the ability of the \gls{MCA} representation to describe the exact dynamics well.
We employ the Full Multiple Spawning approach to evolve the Gaussian basis functions~\cite{Curchod:2020/Book} with classically evolving Gaussian centers~\cite{Ibele:2020/mp/8}.

The rest of the paper is organized as follows.
The theory associated to the adiabatic and \gls{MCA} representations combined with a Gaussian basis for the nuclei is described in Sec.~\ref{sec:theory}.
The computational details are given in Sec.~\ref{sec:comp_detail}.
Section~\ref{sec:results} contains the results and the discussion.
The last section, Sec.~\ref{sec:concl}, concludes this paper.

%%%%%%%%%%%%%%%%%%%%%%%%%%%%%%%%%%%%%%%%%%%%%%%%%%%%%%%%%%%%%%%%%%%%%%%%%%%%%%%%%%
\section{Theory}
\label{sec:theory}

\subsection{The general adiabatic approach for Gaussian expansions}
\label{subsec:Adiab}

The adiabatic representation is defined by its set of electronic adiabatic states $\{\ket{ \varphi_s (\mathbf{X}) };s=0,1,\dots\}$, which are eigenstates of the electronic Hamiltonian $\hat H_e(\mathbf{X})$:
\begin{eqnarray}\label{eq:elecSE}
\hat H_e(\mathbf{X}) \ket{ \varphi_s (\mathbf{X}) } & = & \varepsilon_s (\mathbf{X}) \ket{ \varphi_s (\mathbf{X}) },
\end{eqnarray}
where $\{\varepsilon_s (\mathbf{X})\}$ are the adiabatic potential energy surfaces.
Here $\mathbf{x}$ represents the mass-weighted nuclear coordinates and the hat indicates an abstract operator in the Hilbert space of the electrons.
Projecting the molecular wavefunction $\ket{\Psi (\mathbf{X},t)}$ on this basis gives the Born-Oppenheimer expansion~\cite{Cederbaum:2004/Book}, which is the starting point of many applications in chemistry:
\begin{eqnarray}\label{eq:BHexp}
\ket{\Psi (\mathbf{X},t)} & = & \sum_s \ket{ \varphi_s (\mathbf{X}) } \chi_s (\mathbf{X},t),
\end{eqnarray}
where $\chi_s (\mathbf{X},t)$ is the nuclear wavepacket that evolves on the $s^\text{th}$ electronic adiabatic state.
In the rest of the paper, we assume atomic units and, unless stated otherwise, we drop the nuclear coordinate dependence for the sake of clarity.

The nuclear component are then obtained as a solution of the time-dependent Schr\"odinger with the Hamiltonian given by~\cite{Cederbaum:2004/Book}
\begin{eqnarray}\label{eq:HDV}
H^{DV}_{ss'} 
& = & \bigg[ 
- \frac{1}{2} \boldsymbol{\nabla}^t \boldsymbol{\nabla} 
+ \varepsilon_s  \bigg] \delta_{ss'} 
- \frac{1}{2} \bigg[ 
  \boldsymbol{\tau}_{ss'}^t \boldsymbol{\nabla} 
+ \boldsymbol{\nabla}^t \boldsymbol{\tau}_{ss'} 
+ \sum_{r} \boldsymbol{\tau}_{sr}^t \boldsymbol{\tau}_{rs'} 
\bigg].
\end{eqnarray}
where $\boldsymbol{\nabla}$ is the gradient associated to the mass-weighted nuclear coordinates, $\boldsymbol{\tau}_{ss'}=\braket{\varphi_s}{\boldsymbol{\nabla}\varphi_{s'}}$ is the non-adiabatic coupling vector, and we used bold face letters for vectors and matrices with the superscript '$t$' to indicate transposition.
For the numerical solution, we expand $\chi_s(t)$ as a linear combination of $N_s$ Gaussian functions $\{g(\mathbf{X};\mathbf{Q}_{sk}(t),\mathbf{P}_{sk}(t));k=1,\dots,N_s\}$ on state $s$ where $\mathbf{Q}_{sk}(t)$ and $\mathbf{P}_{sk}(t)$ are time-dependent parameters encoding the centers and the momenta of the Gaussians respectively.
The total wavefunction should now read
\begin{eqnarray}\label{eq:WFDV}
\ket{\Psi (t)} & = & \sum_{s} \sum^{N_s}_{k} C_{sk} (t) \ket{ \varphi_s } g(\mathbf{Q}_{sk}(t),\mathbf{P}_{sk}(t)),
\end{eqnarray}
where $\{C_{sk}(t)\}$ are expansion coefficients that depend only on time.
However, for molecules in vacuum, the electronic Hamiltonian is real, and the adiabatic states are generally chosen to be real too.
In this case, adiabatic electronic states acquire a \gls{GP} when encircling \glspl{CI}, which makes them double-valued with respect to revolving around \glspl{CI} in the nuclear coordinates space.
Since the molecular wavefunction is single-valued, it is clear from Eq.~\ref{eq:BHexp} that, if $\ket{ \varphi_s }$ has double-valued boundary conditions, then the nuclear components are double-valued too since $\chi_s (t)=\braket{ \varphi_s }{\Psi (t)}$.
But linear combination using Gaussian functions cannot exhibit these double-valued boundary conditions, because Gaussian functions are single-valued by construction.
One way around this difficulty is to compensate the double-valuedness by attaching a double-valued phase factor $\mathrm{e}^{\imag\alpha_s}$ that depends only on $\mathbf{X}$ to the adiabatic nuclear components~\cite{Mead:1979/jcp/2284}.
Then, $\mathrm{e}^{-\imag\alpha_s}\chi_s (t)$ is also single-valued and can be expanded using Gaussian functions.
With this transformation, the molecular Hamiltonian for these single-valued nuclear components takes the following form
\begin{eqnarray}\label{eq:HSV}
H^{SV}_{ss'} & = & \mathrm{e}^{-\imag\alpha_{s}} H^{DV}_{ss'} \mathrm{e}^{\imag\alpha_{s'}} \nonumber\\
& = & \bigg[ 
- \frac{1}{2} \boldsymbol{\nabla}^t \boldsymbol{\nabla} 
+ \varepsilon_s  \bigg] \delta_{ss'} 
- \frac{1}{2} \bigg[ 
  \boldsymbol{\tau}_{ss'}^t \boldsymbol{\nabla} 
+ \boldsymbol{\nabla}^t \boldsymbol{\tau}_{ss'} 
+ \sum_{r} \boldsymbol{\tau}_{sr}^t \boldsymbol{\tau}_{rs'} 
\bigg] \mathrm{e}^{\imag(\alpha_{s'}-\alpha_{s})} 
\nonumber\\&&
- \frac{1}{2} \bigg[ 
  \boldsymbol{\upsilon}_s^t \boldsymbol{\upsilon}_s 
+ \boldsymbol{\nabla}^t \boldsymbol{\upsilon}_s 
+ \boldsymbol{\upsilon}_s^t \boldsymbol{\nabla} 
\bigg] \delta_{ss'} 
- \frac{1}{2} \bigg[ 
  \boldsymbol{\tau}_{ss'}^t \boldsymbol{\upsilon}_{s'} 
+ \boldsymbol{\upsilon}_s^t \boldsymbol{\tau}_{ss'} 
\bigg] \mathrm{e}^{\imag(\alpha_{s'}-\alpha_{s})}, 
\end{eqnarray}
where $\boldsymbol{\upsilon}_s=(\imag \boldsymbol{\nabla} \alpha_{s})$.
%In Eq.~\ref{eq:HSV} the gradient can operate beyond the square brackets but not beyond parenthesis.
This Hamiltonian can be compared to the more familiar one in the double-valued real adiabatic basis, $\mathbf{H}^{DV}$, where all phases $\{\alpha_s\}$ can be considered constant and equal.
Unfortunately, the functional form of the phases $\{\alpha_s\}$ is not generally known for real systems.
In this case, the double-valuedness of the nuclear components $\chi_s$ is often overlooked and $\mathbf{H}^{DV}$ is utilized even if the nuclear wavepackets are expanded on a Gaussian basis.

The second difficulty comes from the \gls{DBOC}-related term $\boldsymbol{\tau}_{sr}^t \boldsymbol{\tau}_{rs}$ when the states $r$ and $s$ are degenerate.
These terms can be expressed as 
\begin{eqnarray}\label{eq:HDV}
\boldsymbol{\tau}_{sr}^t \boldsymbol{\tau}_{rs} 
& = & -\frac{ \vert \langle\varphi_{s}\vert\boldsymbol{\nabla}\hat H_e\vert\varphi_{r}\rangle \vert^2 }{(\varepsilon_r-\varepsilon_s)^2},
\end{eqnarray}
which shows that they become singular at \glspl{CI}.
These singularities are known to be non-integrable for basis functions having non-zero populations at \glspl{CI}~\cite{Meek:2016/jcp/184109}.
This non-integrability can be shown for Gaussian functions using a local expansion of the molecular Hamiltonian in the vicinity of a \gls{CI}.
Indeed, when the problematic terms are integrated with Gaussian functions over the full nuclear coordinate space, they give rise to a logarithmic divergence as exposed in App.~\ref{sec:appDBOC}.
To allow for integration, we add a regularization parameter by substituting the square of the energy difference $(\varepsilon_r-\varepsilon_s)^2$ by $(\varepsilon_r-\varepsilon_s)^2+\eta$ where $\eta$ is a small real positive number (see App.~\ref{sec:appDBOC} for more details).

\subsection{The moving crude adiabatic representation}

The situation is different in the \gls{MCA} representation. The electronic states are no more adiabatic but instead solve the electronic Schr\"odinger equation only at one nuclear geometry, which is chosen to be the centers of the Gaussian functions $\{\mathbf{Q}_{sk}(t)\}$:
\begin{eqnarray}\label{eq:MCASE}
\hat H_e(\mathbf{Q}_{sk}(t)) \ket{ \varphi_{s} (\mathbf{Q}_{sk}(t)) } & = & \varepsilon_{s} (\mathbf{Q}_{sk}(t)) \ket{ \varphi_{s} (\mathbf{Q}_{sk}(t)) }.
\end{eqnarray}
The molecular wavefunction now reads
\begin{eqnarray}\label{eq:WFMCA}
\ket{\Psi (t)} & = & \sum_{s} \sum^{N_s}_{k} C_{sk} (t) \ket{ \varphi_s (\mathbf{Q}_{sk} (t)) } g(\mathbf{Q}_{sk}(t),\mathbf{P}_{sk}(t)),
\end{eqnarray}
and the molecular Hamiltonian
\begin{eqnarray}\label{eq:HMCA}
H^{MCA}_{ss',kl} & = & -\frac{1}{2}\braket{ \varphi_{s} (\mathbf{Q}_{sk}(t)) }{ \varphi_{s'} (\mathbf{Q}_{s'l}(t)) } \boldsymbol{\nabla}^t \boldsymbol{\nabla} + \braOket{ \varphi_{s} (\mathbf{Q}_{sk}(t)) }{ {\hat H}_e }{ \varphi_{s'} (\mathbf{Q}_{s'l}(t)) }.
\end{eqnarray}
This last Hamiltonian is different from the Hamiltonians using adiabatic states (Eqs.~(\ref{eq:HSV}-\ref{eq:HDV})) in the sense there are no non-adiabatic couplings $\boldsymbol{\tau}_{ss'}$ involved.
This is due to the fact that \gls{MCA} states do not depend on the nuclear coordinates and are therefore strictly diabatic.
Therefore, there are no terms in the Hamiltonian that are non-integrable with a Gaussian basis.
Another strong advantage is that, since the \gls{MCA} states do not depend on the nuclear coordinates, they are single-valued in the nuclear coordinates space by construction.
Hence, Gaussian functions can be employed for the expansion of the nuclear wavepackets.

%%%%%%%%%%%%%%%%%%%%%%%%%%%%%%%%%%%%%%%%%%%%%%%%%%%%%%%%%%%%%%%%%%%%%%%%%%%%%%%%%%
\section{Computational details}
\label{sec:comp_detail}

\subsection{Hamiltonian models}

We utilize 2-dimensional 2-state linear vibronic coupling models to simulate the electron transfers in \gls{BMA} and in the butatriene cation.
The general form of the model is given in the diabatic representation and in mass and frequency weighted coordinates
\begin{eqnarray}\label{eq:LVC}
\mathbf{H}^{dia} & = & -\frac{1}{2} \boldsymbol{\nabla}^t\boldsymbol{\omega}\boldsymbol{\nabla} \left(\begin{smallmatrix}1&0\\0&1\end{smallmatrix}\right) + \mathbf{H}_e, \\\label{eq:LVCHe}
\mathbf{H}_e & = & \frac{1}{2}\left[ (\mathbf{X}^t\boldsymbol{\omega}\mathbf{X}+\mathbf{X}^t\boldsymbol{\sigma}+\epsilon_\sigma) \left(\begin{smallmatrix}1&0\\0&1\end{smallmatrix}\right) + (\mathbf{X}^t\boldsymbol{\kappa}+\epsilon_\kappa) \left(\begin{smallmatrix}-1&0\\0&1\end{smallmatrix}\right) + (\mathbf{X}^t\boldsymbol{\lambda}+\epsilon_\lambda) \left(\begin{smallmatrix}0&1\\1&0\end{smallmatrix}\right) \right].
\end{eqnarray}
The models are obtained from Ref.~\citenum{Ryabinkin:2014/jcp/214116}, and originally from Refs.~\citenum{Cattarius:2001/jcp/2088,Izmaylov:2011/jcp/234106}, after transformation to mass and frequency weighted coordinates.
The parameters are given in Tab.~\ref{tab:param}
\begin{table}[!h]
\caption{Parameters for the two molecular systems in atomic units. Parameters come from Ref.~\citenum{Ryabinkin:2014/jcp/214116} after transformation to mass and frequency weighted coordinates.}%%%Table caption goes here
\label{tab:param}
\centering
\begin{tabular}{lllllll}%%%The number of columns has to be defined here
\hline
System    & $\omega_{11}$     & $\omega_{22}$     & $\sigma_1$         & $\kappa_1$          & $\lambda_2$       & $\epsilon_\kappa$ \\
\hline
C$_4$H${_4}^{+}$&$9.56\cdot 10^{-3}$&$3.35\cdot 10^{-3}$&$-1.88\cdot 10^{-2}$&$1.88\cdot 10^{-2}$  &$2.12\cdot 10^{-2}$&$1.45\cdot 10^{-3}$\\
\gls{BMA} &$7.74\cdot 10^{-3}$&$6.68\cdot 10^{-3}$&$0.00$              &$-2.126\cdot 10^{-2}$&$9.90\cdot 10^{-4}$&$0.00$ \\
\hline
\end{tabular}
\begin{tabular}{ll}%%%The number of columns has to be defined here
\hline
For both systems & $\omega_{12}=\omega_{21}=0$, $\sigma_2=\kappa_2=\lambda_1=0$, $\epsilon_\lambda=\epsilon_\sigma=0$ \\
\hline
\end{tabular}

\vspace*{-4pt}
\end{table}%%%End of the table

%For this model, the phases $\{\alpha_s\}$ in Eq.~(\ref{eq:HSV}) takes the form~\cite{}
%\begin{eqnarray}\label{eq:alpha}
%\alpha_1 = \alpha_2 & = & \frac{1}{2}\arctan2\left( \mathbf{X}^t\boldsymbol{\lambda}+\epsilon_\lambda , \mathbf{X}^t\boldsymbol{\kappa}+\epsilon_\kappa \right),
%\end{eqnarray}
%where $\arctan2$ is the two-argument arctangent function.
%Using this form of the phase results in the fact that $\boldsymbol{\upsilon}_s=\boldsymbol{\tau}_{12}$.
%Therefore, the term $\boldsymbol{\upsilon}_s^t \boldsymbol{\upsilon}_s$ is also non-integrable.

\subsection{The Gaussian basis}

The Gaussian basis is chosen as frozen-width coherent states with the form
\begin{eqnarray}\label{eq:gdef}
g ( \mathbf{Q}_{sk}(t),\mathbf{P}_{sk}(t) ) 
& = & \exp\left( -\frac{1}{2} \mathbf{X}^t\mathbf{X} + \mathbf{X}^t\mathbf{b}_{sk} (t) + c_{sk} (t) \right),
\end{eqnarray}
where
\begin{eqnarray}\label{eq:TDGaussparam}
\mathbf{b}_{sk} (t) & = & \mathbf{Q}_{sk}(t) + \imag\mathbf{P}_{sk}(t),\\
c_{sk} (t) & = & -\frac{1}{2} \mathbf{Q}_{sk}^t(t) \mathbf{b}_{sk} (t) - \frac{\mathcal{D}\log(\pi)}{4},
\end{eqnarray}
and $\mathcal{D}$ is the dimensionality of the system.

The time evolution of the parameters is dictated by classical equations of motion on the corresponding adiabatic surfaces:
\begin{eqnarray}\label{eq:gdef}
\dot{\mathbf{Q}}_{sk}(t) 
& = & \boldsymbol{\omega}\mathbf{P}_{sk} (t) , \\
\dot{\mathbf{P}}_{sk}(t) 
%& = & -\boldsymbol{\nabla}_Q \varepsilon_s (\mathbf{Q}_{sk}(t)) \nonumber\\
%& = & -\frac{1}{2}\boldsymbol{\nabla}_Q \left( (\mathbf{X}^t\boldsymbol{\omega}\mathbf{X}+\mathbf{X}^t\boldsymbol{\sigma}+\epsilon_\sigma) \pm \sqrt{ (\mathbf{X}^t\boldsymbol{\kappa}+\epsilon_\kappa)^2 + (\mathbf{X}^t\boldsymbol{\lambda}+\epsilon_\lambda)^2 } \right) \nonumber\\
& = & -\frac{1}{2} \left( 2\boldsymbol{\omega}\mathbf{Q}_{sk}(t)+\boldsymbol{\sigma} \pm \frac{\boldsymbol{\kappa}(\mathbf{Q}^t_{sk}(t)\boldsymbol{\kappa}+\epsilon_\kappa) + \boldsymbol{\lambda}(\mathbf{Q}^t_{sk}(t)\boldsymbol{\lambda}+\epsilon_\lambda)}{\sqrt{ (\mathbf{Q}^t_{sk}(t)\boldsymbol{\kappa}+\epsilon_\kappa)^2 + (\mathbf{Q}^t_{sk}(t)\boldsymbol{\lambda}+\epsilon_\lambda)^2 }} \right),
\end{eqnarray}
where the plus (minus) sign is associated to the upper (lower) adiabatic state.

The spawning procedure follows the prescription of Ref.~\citenum{Curchod:2020/Book}.
The technical details of this spawning procedure as well as the integration scheme for equations of motion and the integration over the Gaussian basis of the Hamiltonian is described in the supplementary material~\cite{suppmat}.
Gaussian-based calculations are compared to exact calculations using a finite basis representation of the Hamiltonians in Eqs.~(\ref{eq:LVC}-\ref{eq:LVCHe}), which is also described in the supplementary material~\cite{suppmat}.
All simulations are done with the Octave package~\cite{octave}.

%%%%%%%%%%%%%%%%%%%%%%%%%%%%%%%%%%%%%%%%%%%%%%%%%%%%%%%%%%%%%%%%%%%%%%%%%%%%%%%%%%
\section{Results and discussion}
\label{sec:results}

We test the methods on the internal conversion process.
The initial state is depicted by a single Gaussian basis function placed on the upper electronic state at positions $(0.632,0)$ for the butatriene cation and $(-1.366,0)$ for \gls{BMA} with zero momentum in both cases.
We then generate other unpopulated Gaussians according to this initial Gaussian distribution with symmetric sampling.
The starting number of Gaussian functions $N_i$ is given in Tab.~\ref{tab:Ng}
Along the dynamics, new spawned Gaussian basis functions increase this number to a final one $N_f$.
We start by testing the simulations in adiabatic representation, and then test the \gls{MCA} representation.
These tests are done by comparing the adiabatic populations with respect to the exact simulations.

\begin{table}[!h]
\caption{Initial and final number of basis functions used in the Gaussian-based simulations for both models in adiabatic and \gls{MCA} representations.}%%%Table caption goes here
\label{tab:Ng}
\centering
\begin{tabular}{lcc}%%%The number of columns has to be defined here
\hline
          &       \gls{BMA}      &   C$_4$H$_4^{+}$      \\
\hline
Adiabatic & $N_i=256$; $N_f=328$ & $N_i=346$; $N_f=416$  \\
\gls{MCA} & $N_i=251$; $N_f=326$ & $N_i=339$; $N_f=442$  \\
\hline
\end{tabular}

\vspace*{-4pt}
\end{table}%%%End of the table

\subsection{Adiabatic simulations}

We follow the adiabatic population obtained using a Gaussian basis in combination with the double-valued electronic states.
This is equivalent to employ the Hamiltonian given in Eq.~\ref{eq:HDV} but overlooking the \gls{GP} since the Gaussian basis is single valued.

\begin{figure}[!h]
\centering
\includegraphics[width=0.5\textwidth]{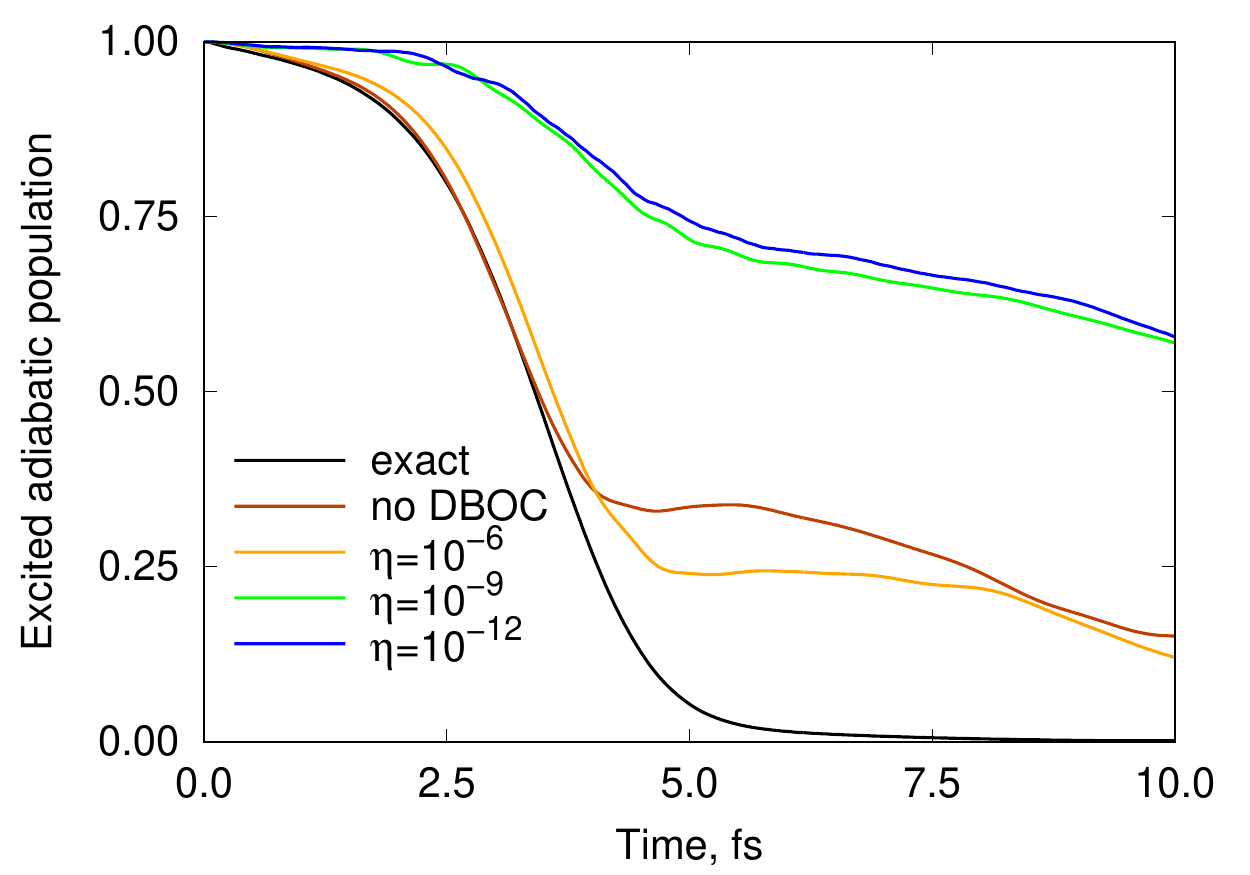}
\caption{Comparison between adiabatic simulation without \gls{GP} for various values of the regularization parameter $\eta$, for \gls{BMA}.}
\label{fig:Adiab_BMA}
\end{figure}

We first test the impact of \gls{DBOC} in the dynamics of \gls{BMA} where it plays an important role due to the small diabatic coupling~\cite{Ryabinkin:2014/jcp/214116}.
In this case, \gls{DBOC} plays the role of an infinite potential at the \gls{CI} that prevents the nuclear wavepackets to access the region of strong non-adiabatic couplings and thus diminishes the non-adiabatic transitions.
This effect is however counterbalanced by a similar term generated by the \gls{GP} such that the wavepackets can approach the \gls{CI}, which allow for larger non-adiabatic transitions.
Hence, a calculation that properly includes the \gls{DBOC} term without \gls{GP} must show a decrease in the population transfer as observed in Ref.~\citenum{Ryabinkin:2014/jcp/214116}.
This will be our test to assess the proper account for \gls{DBOC}.
Since, the \gls{DBOC} terms, $\boldsymbol{\tau}_{ss'}^t \boldsymbol{\tau}_{s's}$, are non-integrable, we employ a regularized version of these terms as depicted at the end of Sec.~\ref{sec:theory}-\ref{subsec:Adiab} (and also developed in App.~\ref{sec:appDBOC}) with various values of $\eta$.
Results are given in Fig.~\ref{fig:Adiab_BMA}.
They show that for large values of $\eta$, the repelling effect of \gls{DBOC} is significantly reduced and does not play a significant role.
However, for small values, $\eta=10^{-9},10^{-12}$, \gls{DBOC} clearly diminishes the population transfer in agreement with Ref.~\citenum{Ryabinkin:2014/jcp/214116}.
We deduce that if $\eta$ is sufficiently small, the regularized \gls{DBOC} can reproduce the effect of the exact \gls{DBOC}.
We then test the compensation of \gls{GP} and \gls{DBOC}: it was previously found that when \gls{GP} is not included in the simulation, not including \gls{DBOC} improves the results~\cite{Ryabinkin:2014/jcp/214116}.
This is indeed what we observe in Fig.~\ref{fig:Adiab_BMA} for the `no DBOC' case, where the initial population dynamics is captured when both \gls{GP} and \gls{DBOC} are overlooked.

\begin{figure}[!h]
\centering
\begin{tabular}{lclc}
a)\vspace{-0.2cm}&&b)\vspace{-0.2cm}&\\
&\hspace{-0.7cm}\includegraphics[width=0.5\textwidth]{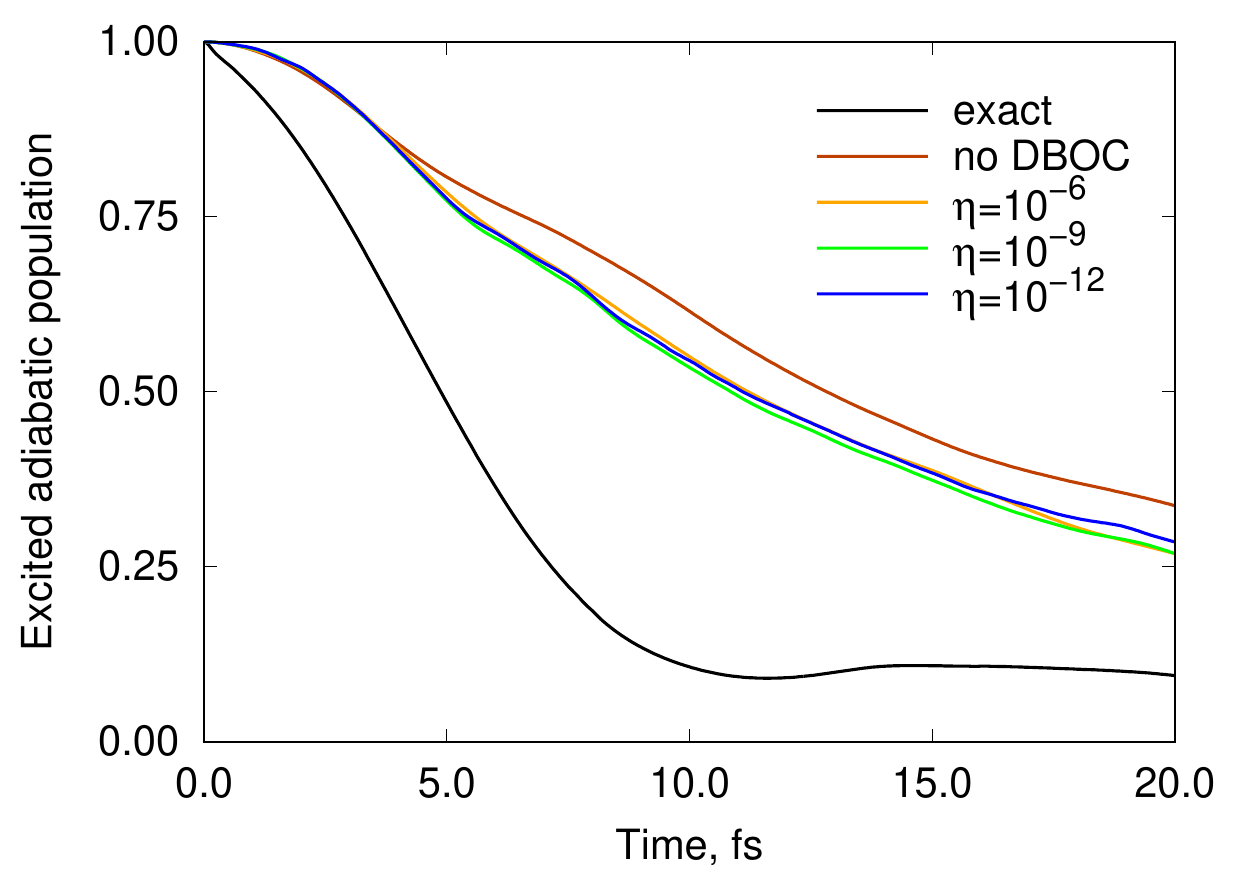}&&\hspace{-0.7cm}\includegraphics[width=0.5\textwidth]{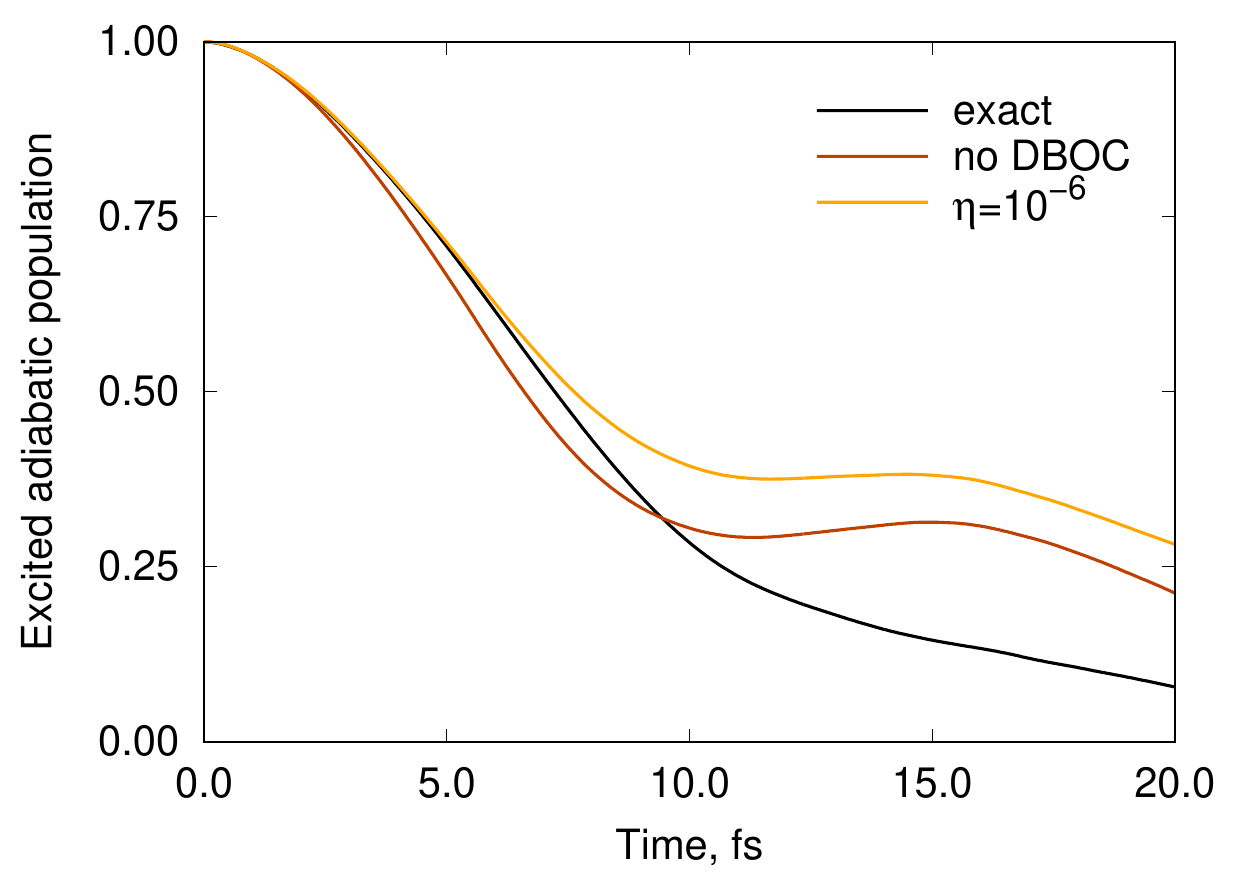}
\end{tabular}
\caption{Comparison between adiabatic simulations without \gls{GP} for various values of the regularization parameter $\eta$, for the butatriene cation. Initial conditions are taken as: a) one Gaussian function centered at $(-1.366,0)$, and b) two Gaussian functions at $(-1.366,\pm 2.8\cdot 10^{-3})$ with opposite coefficients, on the upper adiabatic state in both cases.}
\label{fig:Adiab_C4H4}
\end{figure}

We now test the impact of \gls{GP} in enhancing the transfer probability for the components of the wavepackets with zero angular momentum with respect to the \gls{CI}~\cite{Ryabinkin:2014/jcp/214116}.
This effect is better illustrated in the case of the butatriene cation, where the wavepacket has a large component ($87\%$) with zero angular momentum when arriving at the \gls{CI}~\cite{Ryabinkin:2014/jcp/214116}.
Figure~\ref{fig:Adiab_C4H4}-a) shows that \gls{DBOC} only has a minor impact on the population evolution.
Indeed, including \gls{DBOC} or not, does not significantly affect the population dynamics.
This absence of effect also appears through the fact that the value of the regularization parameter does not significantly change the results.
To expose the \gls{GP} effect associated to enhancing probability transfer, the component of the wavepacket that is associated to a zero expectation value of the angular momentum is artificially reduced by changing the initial state as in Ref.~\citenum{Ryabinkin:2014/jcp/214116}.
We use an initial state defined as a linear combination of two Gaussians centered at $(-1.366,\pm 2.8\cdot 10^{-3})$ with coefficients of opposite signs $C_1(0)=1=-C_2(0)$.
This initial state reproduces the original single-Gaussian state multiplied by $X_2$, which creates a nodal line $X_2=0$ in the wave-
packet and eliminates the component associated to a zero expectation value of the angular momentum.
The result, given in Fig.~\ref{fig:Adiab_C4H4}-b), clearly shows that employing this new initial condition reduces the difference between the exact population dynamics and the one obtained in the Gaussian basis (without \gls{GP}).
Consequently, the Gaussian-based method now captures the non-adiabatic transition.

We observe that integration of \gls{DBOC} can be mimicked with a right choice of the regularization parameter $\eta$.
Hence, non-integrability of \gls{DBOC} appears to be less problematic than expected for the models studied in this article.
We also observed that all the results in adiabatic representation agree with the previously reported simulations using the same models~\cite{Ryabinkin:2014/jcp/214116}, which are not including the \gls{GP}.
This numerical argument shows that \gls{GP} is not included when using a Gaussian basis with the Hamiltonian given by Eq.~\ref{eq:HDV}.

\subsection{\gls{MCA} simulations}

\begin{figure}[!h]
\centering
\begin{tabular}{lclc}
a)\vspace{-0.2cm}&&b)\vspace{-0.2cm}&\\
&\hspace{-0.7cm}\includegraphics[width=0.5\textwidth]{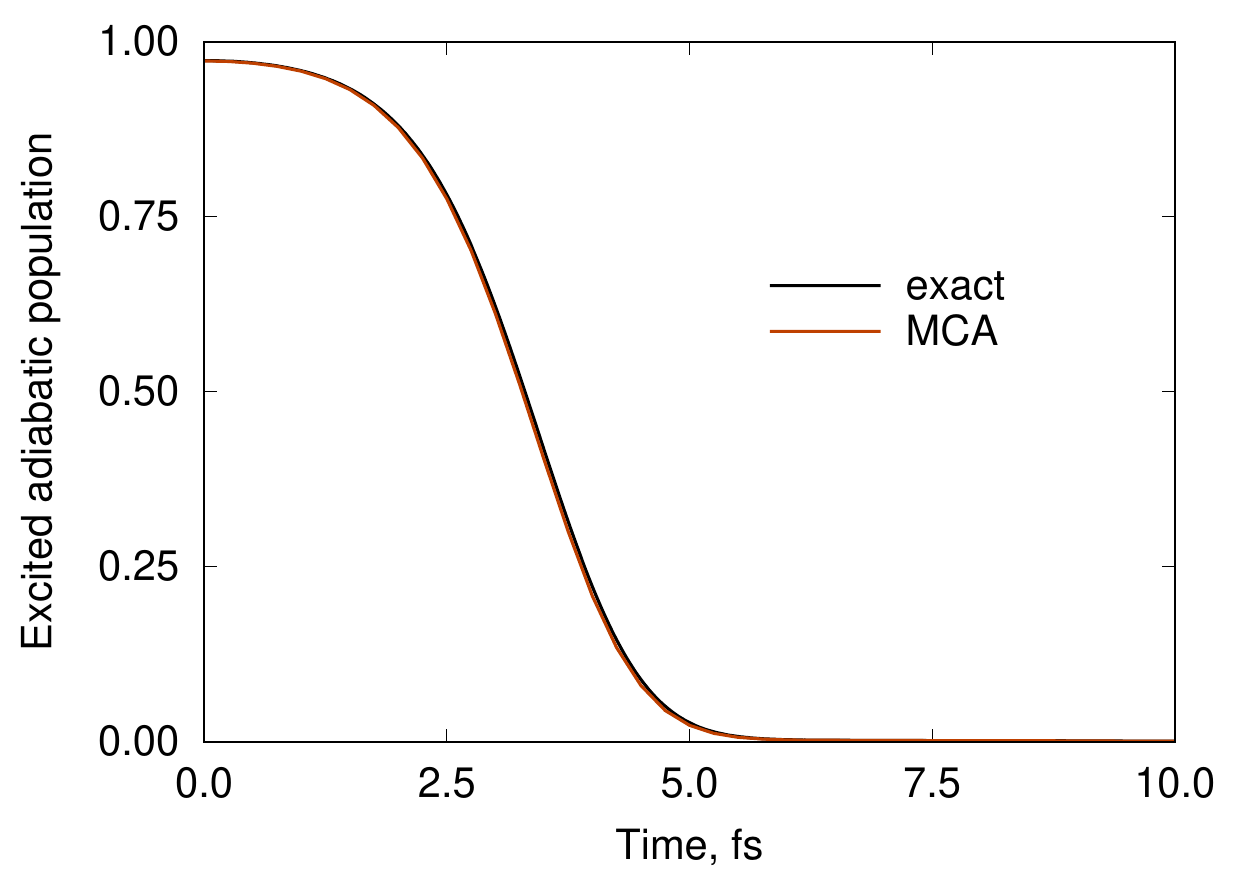}&&\hspace{-0.7cm}\includegraphics[width=0.5\textwidth]{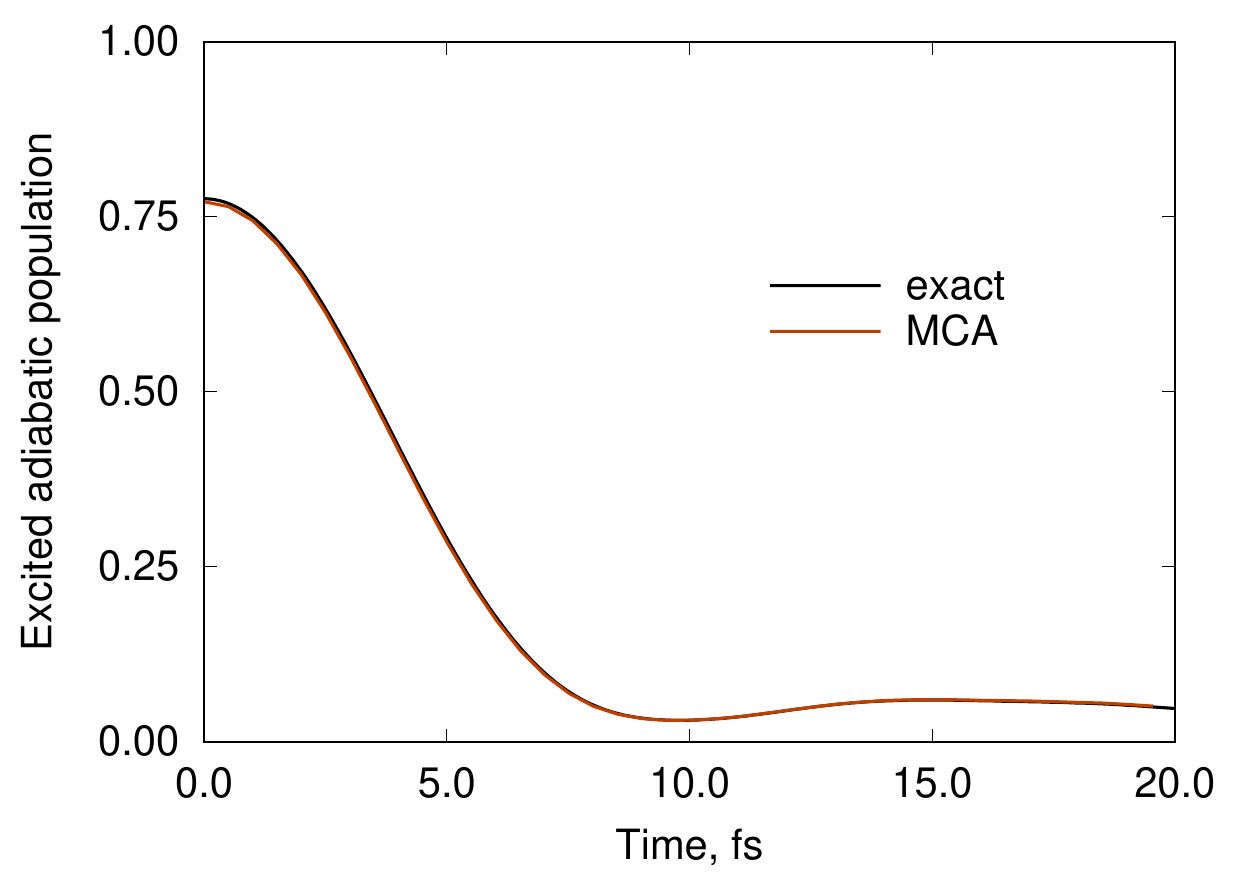}
\end{tabular}
\caption{Comparison between the upper adiabatic state time-dependent population using the \gls{MCA} representation and the exact calculation for: a) \gls{BMA} and b) C$_4$H${_4}^{+}$.}
\label{fig:MCA}
\end{figure}

In this case, the initial state population is not fully on the upper adiabatic state because the \gls{MCA} electronic states are not adiabatic.
Indeed, creating an initial state that is strictly on the upper adiabatic state would require a linear combination of \gls{MCA} states whose positions are optimized to maximize overlap with the initial Gaussian state in the adiabatic representation.
However, such a construction is not necessary because we are interested in showing that dynamics in the \gls{MCA} states converges to the exact dynamics independently from the initial conditions.
The adiabatic population can still be calculated from the \gls{MCA} calculations by using the scheme explained in the supplementary material~\cite{suppmat}.
For \gls{BMA}, the initial populated Gaussian is far enough from the \gls{CI} and the diabatic coupling is small enough so that the initial upper state population is close to $1$.
However, for the butatriene cation, the initial upper population is closer to $0.75$.
We model this same initial states in the exact calculations as well.
The upper state adiabatic population is then calculated using the Gaussian integration technique described in the supplementary material~\cite{suppmat}. 
The comparison between the exact population evolution and the dynamics obtained using the \gls{MCA} representation is given in Fig.~\ref{fig:MCA}.
These results show that the \gls{MCA} representation reproduces the exact probability transfer.

\begin{figure}[!h]
\centering
\begin{tabular}{lclc}
a)\vspace{-0.2cm}&&b)\vspace{-0.2cm}&\\
%&\hspace{-0.7cm}\includegraphics[width=0.5\textwidth]{auto_BMA}&&\hspace{-0.7cm}\includegraphics[width=0.5\textwidth]{auto_C4H4}
&\hspace{-0.7cm}\includegraphics[width=0.5\textwidth]{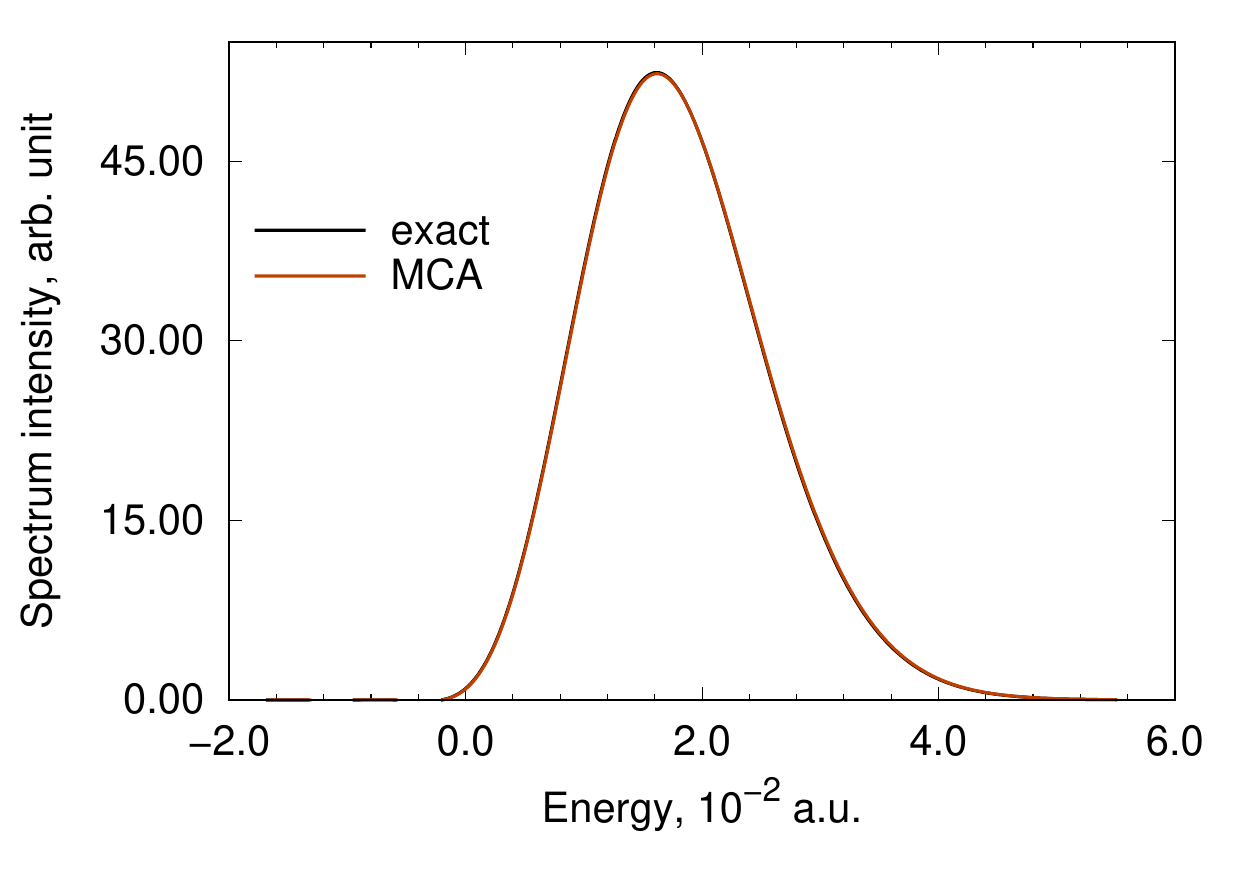}&&\hspace{-0.7cm}\includegraphics[width=0.5\textwidth]{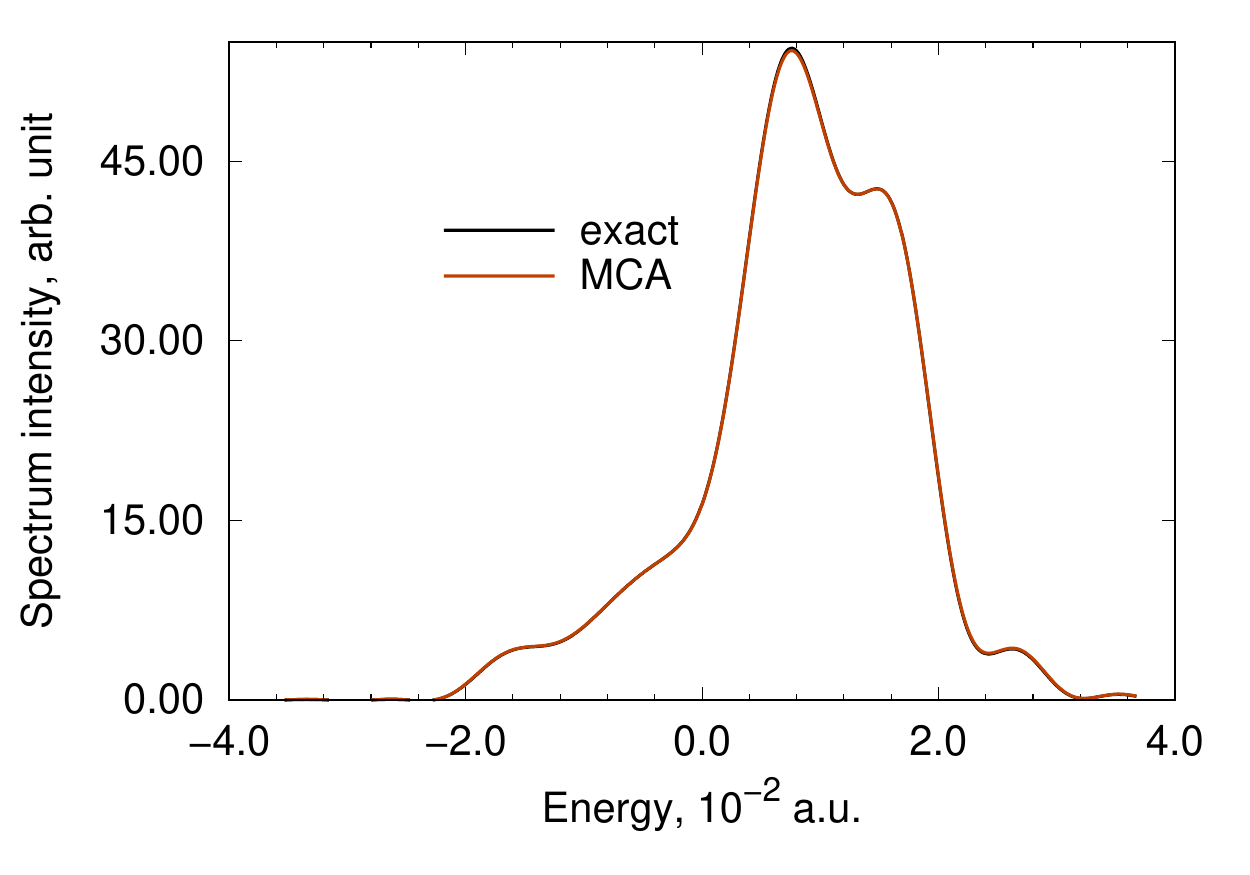}
\end{tabular}
\caption{Comparison between the spectra obtained using the \gls{MCA} representation and the exact calculation for: a) \gls{BMA} and b) C$_4$H${_4}^{+}$.}
\label{fig:spec}
\end{figure}

We test the convergence by comparing the autocorrelation functions $\vert\langle\Psi(0)\vert\Psi(t)\rangle\vert$ for these dynamics after extracting the spectra calculated by Fourier transform of these autocorrelation functions.
These spectra, given in Fig.~\ref{fig:spec}, show perfect agreement between the spetra obtain with the exact calculation and the one using the \gls{MCA} representation.

Both the adiabatic population evolution and the Fourier transform of the autocorrelation function agree to show that the \gls{MCA} representation can be used to generate the exact dynamics.

%%%%%%%%%%%%%%%%%%%%%%%%%%%%%%%%%%%%%%%%%%%%%%%%%%%%%%%%%%%%%%%%%%%%%%%%%%%%%%%%%%
\section{Conclusion}
\label{sec:concl}

In this paper, we assess the ability of Gaussian-based methods to reproduce the exact dynamics while employing adiabatic electronic states.

We observed that using Gaussian functions in the adiabatic representation exhibits the same problems as in Ref.~\citenum{Ryabinkin:2014/jcp/214116}: not including \gls{GP} results in time evolutions that deviate from the exact solution.
Therefore, we can conclude that using Gaussian functions fails in properly accounting \gls{GP}, and one must be cautious when the dynamics involve \glspl{CI}.

We also studied the problem of the non-integrability of \gls{DBOC} in the adiabatic representation.
As expected from Ref.~\citenum{Meek:2016/jcp/184109}, we found that this term is indeed non-integrable in a Gaussian basis.
Nevertheless, we applied a regularization that makes the integral calculation possible thanks to an extra small parameter in the denominator.
We observed that for small enough values of this parameter, the dynamics behaves similarly to what was observed in Ref.~\citenum{Ryabinkin:2014/jcp/214116}.
Hence, non-integrability of \gls{DBOC} is not a major problem for the dynamical processes that are studied here.

Finally, we tested the dynamics obtained using the \gls{MCA} representation.
We found that this representation can properly reproduce the exact time evolution of the adiabatic population, and also the autocorrelation function.
Thus, we can conclude that the \gls{MCA} representation is capable of describing correctly internal conversion processes.

These numerical evidences of \gls{MCA} validity are important since this representation still employs the adiabatic states (at the Gaussians' center) obtained from quantum chemistry packages, but with the strong advantage that it does not need any integral regularization or extra phase factor.
Furthermore, the \gls{MCA} representation also has the advantage that it alleviates the need to potential energy models by applying Gaussian integration techniques on the exact molecular Hamiltonian~\cite{Joubert:2018/jcp/114102}.
Hence, validating \gls{MCA} in the context of non-adiabatic transitions is an important step before further developments related to the exact integration of the molecular Hamiltonian.

%%%%%%%%%%%%%%%%%%%%%%%%%%%%%%%%%%%%%%%%%%%%%%%%%%%%%%%%%%%%%%%%%%%%%%%%%%%%%%%%%%
\section*{ACKNOWLEDGEMENTS}
LJD and RM are grateful to Michael Schuurman for fruitful discussions on the implementation of the full multiple spawning algorithm, and Artur Izmaylov, Mina Asaad, and Alexander Mitrushchenkov for their comments on the manuscript.
This work was supported by the 2021 Mobility and International Cooperation Incentive Call of the I-SITE FUTURE.

\appendix

%%%%%%%%%%%%%%%%%%%%%%%%%%%%%%%%%%%%%%%%%%%%%%%%%%%%%%%%%%%%%%%%%%%%%%%%%%%%%%%%%%
\section*{Exposition and regularization of the diverging term in \gls{DBOC}}
\label{sec:appDBOC}

In this appendix, we derive an expression of the \gls{DBOC} integrated over Gaussian basis functions for the models of the form given in Eq.~\ref{eq:LVC}.
The non-adiabatic coupling vector takes the form
\begin{eqnarray}\label{eq:NACV}
\boldsymbol{\tau}_{12} & = & \frac{1}{2} \frac{\boldsymbol{\kappa}(\mathbf{X}^t\boldsymbol{\lambda}+\epsilon_\lambda) - \boldsymbol{\lambda}(\mathbf{X}^t\boldsymbol{\kappa}+\epsilon_\kappa)}{ (\mathbf{X}^t\boldsymbol{\kappa}+\epsilon_\kappa)^2 + (\mathbf{X}^t\boldsymbol{\lambda}+\epsilon_\lambda)^2 }.
\end{eqnarray}
For further simplification of the expressions, we operate a translation to place the \gls{CI}'s position $\mathbf{X}_{CI}$ in the center of the new coordinate system $\tilde{\mathbf{X}}=\mathbf{X}-\mathbf{X}_{CI}$.
The \gls{CI}'s position is obtained by minimizing the energy under the constraints $\mathbf{X}^t\boldsymbol{\kappa}+\epsilon_\kappa=\mathbf{X}^t\boldsymbol{\lambda}+\epsilon_\lambda=0$:
\begin{eqnarray}\label{eq:XCI}
\mathbf{X}_{CI} 
& = & -\frac{1}{2}\boldsymbol{\omega}^{-1}\left(\boldsymbol{\omega}-(\begin{smallmatrix}\boldsymbol{\kappa}&\boldsymbol{\lambda}\end{smallmatrix})\left[(\begin{smallmatrix}\boldsymbol{\kappa}&\boldsymbol{\lambda}\end{smallmatrix})^t\boldsymbol{\omega}^{-1}(\begin{smallmatrix}\boldsymbol{\kappa}&\boldsymbol{\lambda}\end{smallmatrix})\right]^{-1}(\begin{smallmatrix}\boldsymbol{\kappa}&\boldsymbol{\lambda}\end{smallmatrix})^t\right)\boldsymbol{\omega}^{-1}\boldsymbol{\sigma} \nonumber\\
& & -\boldsymbol{\omega}^{-1}(\begin{smallmatrix}\boldsymbol{\kappa}&\boldsymbol{\lambda}\end{smallmatrix})\left[(\begin{smallmatrix}\boldsymbol{\kappa}&\boldsymbol{\lambda}\end{smallmatrix})^t\boldsymbol{\omega}^{-1}(\begin{smallmatrix}\boldsymbol{\kappa}&\boldsymbol{\lambda}\end{smallmatrix})\right]^{-1}(\begin{smallmatrix}\epsilon_\kappa\\\epsilon_\lambda\end{smallmatrix}).
\end{eqnarray}
Furthermore, we rotate the coordinates to diagonalize the matrix $\boldsymbol{\kappa}\boldsymbol{\kappa}^t+\boldsymbol{\lambda}\boldsymbol{\lambda}^t$, using the rotation $\mathbf{R}$ so that $\bar{\mathbf{X}}=\mathbf{R}^t\tilde{\mathbf{X}}$ and $\mathbf{R}^t(\boldsymbol{\kappa}\boldsymbol{\kappa}^t+\boldsymbol{\lambda}\boldsymbol{\lambda}^t)\mathbf{R}=\bar{\mathbf{g}}$ has only two non-zero eigenvalues (by construction) that we order as $\bar{\mathbf{g}}=\mathrm{diag}(\bar{g}_1 ,\bar{g}_2 ,0 ,\dots)$.
These combined translation and rotation redefine several quantities that we indicate with a bar and that are summarized here:
\begin{eqnarray}\label{eq:translDefs}
\bar{\mathbf{X}} & = & \mathbf{R}^t(\mathbf{X}-\mathbf{X}_{CI}), \\
\bar{\boldsymbol{\sigma}} & = & \mathbf{R}^t ( \boldsymbol{\sigma} + 2\boldsymbol{\omega}\mathbf{X}_{CI}), \\
\bar{\epsilon}_\sigma & = & \epsilon_\sigma + \mathbf{X}_{CI}^t\boldsymbol{\sigma} + \mathbf{X}_{CI}^t\boldsymbol{\omega}\mathbf{X}_{CI}, \\
\bar{\epsilon}_\kappa = \bar{\epsilon}_\lambda & = & 0, \\
\bar{\mathbf{g}} & = & \mathbf{R}^t(\boldsymbol{\kappa}\boldsymbol{\kappa}^t+\boldsymbol{\lambda}\boldsymbol{\lambda}^t)\mathbf{R}, \\
\bar{\mathbf{K}} & = & -\mathbf{R}^t( \boldsymbol{\kappa}\boldsymbol{\lambda}^t - \boldsymbol{\lambda}\boldsymbol{\kappa}^t )\boldsymbol{\omega}( \boldsymbol{\kappa}\boldsymbol{\lambda}^t - \boldsymbol{\lambda}\boldsymbol{\kappa}^t )\mathbf{R}, \\
\bar{\mathbf{b}}_{ss'kl} (t) & = & \frac{1}{2}\mathbf{R}^t\left( \mathbf{b}_{sk}^*(t) + \mathbf{b}_{s'l}(t) - 2\mathbf{X}_{CI} \right), \\
\bar{c}_{ss'kl} (t) & = & c_{sk}^*(t) + \mathbf{X}_{CI}^t\mathbf{b}_{sk}^*(t) + c_{s'l}(t) + \mathbf{X}_{CI}^t\mathbf{b}_{s'l}(t) - \mathbf{X}_{CI}^t\mathbf{X}_{CI}.
\end{eqnarray}
The \gls{DBOC} term now takes the form
\begin{eqnarray}\label{eq:DBOC1}
\boldsymbol{\tau}_{12}^t\boldsymbol{\tau}_{21} = \boldsymbol{\tau}_{21}^t\boldsymbol{\tau}_{12} 
& = & -\frac{1}{4} \frac{
\bar{\mathbf{X}}^t\bar{\mathbf{K}}\bar{\mathbf{X}}
}{ ( \bar{\mathbf{X}}^t\bar{\mathbf{g}}\bar{\mathbf{X}} )^2 }.
\end{eqnarray}
Since integration of \gls{DBOC} is expected to diverge, we integrate a regularized version of the operator by adding a small positive number $\eta$ in the denominator function. The integrals we need to evaluate now read
\begin{eqnarray}\label{eq:DBOCint}
I & = & \frac{1}{8} \int_{\mathbb{R}^\mathcal{D}}\!\!\mathrm{d}\tilde{V} 
\frac{
\bar{\mathbf{X}}^t\bar{\mathbf{K}}\bar{\mathbf{X}}
\,\,\mathrm{e}^{ - \bar{\mathbf{X}}^t\bar{\mathbf{X}} + 2\bar{\mathbf{X}}^t\bar{\mathbf{b}}_{ss'kl} (t) + \bar{c}_{ss'kl} (t) }
}{ ( \bar{\mathbf{X}}^t\bar{\mathbf{g}}\bar{\mathbf{X}} + \eta )^2 }.
\end{eqnarray}
For the evaluation, we use the definition of the Gamma function to rewrite the denominator as
\begin{eqnarray}\label{eq:Gamma}
( \bar{\mathbf{X}}^t\bar{\mathbf{g}}\bar{\mathbf{X}} + \eta )^{-2}
& = & 2\int_0^\infty \mathrm{d}r\,r^3\,\mathrm{e}^{-r^2 (\bar{\mathbf{X}}^t\bar{\mathbf{g}}\bar{\mathbf{X}}+\eta)}.
\end{eqnarray}
Substituting the denominator in Eq.~\ref{eq:DBOCint} using the relation Eq.~\ref{eq:Gamma}, an performing the Gaussian integration, we obtain a one-dimensional integral:
\begin{eqnarray}\label{eq:DBOC2}
I 
& = & \frac{S_{ss'kl}(t)}{4} \int_0^\infty \mathrm{d}r\,r^3\,\frac{
\mathrm{e}^{ - r^2 \eta - r^2 \left( \frac{\bar{b}_{ss'kl,1}^2 (t) \bar{g_1}}{1+r^2 \bar{g_1}} + \frac{\bar{b}_{ss'kl,2}^2 (t) \bar{g_2}}{1+r^2 \bar{g_2}} \right) }
}{\sqrt{(1+r^2 \bar{g}_1)(1+r^2 \bar{g}_2)}}\,
\times\Bigg(
  \frac{\bar{b}_{ss'kl,1}^2 (t)\bar{K}_{11}}{(1+r^2 \bar{g}_1)^2}
\nonumber\\&&
+ 2\frac{\bar{b}_{ss'kl,1} (t)\bar{b}_{ss'kl,2} (t)\bar{K}_{12}}{(1+r^2 \bar{g}_1)(1+r^2 \bar{g}_2)}
+ \frac{\bar{b}_{ss'kl,2}^2 (t)\bar{K}_{22}}{(1+r^2 \bar{g}_2)^2}
+ \frac{1}{2} \frac{\bar{K}_{11}}{1+r^2 \bar{g}_1} 
+ \frac{1}{2} \frac{\bar{K}_{22}}{1+r^2 \bar{g}_2} 
\Bigg), 
\end{eqnarray}
where $S_{ss'kl}(t)=\pi^{\mathcal{D}/2} \exp( \bar{\mathbf{b}}_{ss'kl}^t (t)\bar{\mathbf{b}}_{ss'kl} (t) + \bar{c}_{ss'kl} (t) )$ is the Gaussian overlap.
This integral can be brought to an integral over a finite range by performing the change of variable $r^2=u^2/(1-u^2\bar{g}_1)$ [$r^3\mathrm{d} r=u^3\mathrm{d}u/(1-u^2\bar{g}_1)^3$] and defining $\Delta\bar{g}=\bar{g}_1-\bar{g}_2$:
\begin{eqnarray}\label{eq:DBOC3}
I 
%& = & \frac{\pi^{\mathcal{D}/2}}{4} \mathrm{e}^{ \vert\vert\bar{\mathbf{b}}_{ss'kl} (t)\vert\vert^2 + \bar{c}_{ss'kl}^{*} (t) } \int_0^\infty \mathrm{d}r\,r^3\,\frac{
%\mathrm{e}^{ - r^2 \eta - r^2 \left( \frac{\bar{b}_{ss'kl,1}^2 (t) \bar{g_1}}{1+r^2 \bar{g_1}} + \frac{\bar{b}_{ss'kl,2}^2 (t) \bar{g_2}}{1+r^2 \bar{g_2}} \right) }
%
%}{\sqrt{(1+r^2 \bar{g}_1)(1+r^2 \bar{g}_2)}}\,\nonumber\\&&
%\times\left(
%  \frac{\bar{b}_{ss'kl,1}^2 (t)\bar{K}_{11}}{(1+r^2 \bar{g}_1)^2}
%+ \frac{\bar{b}_{ss'kl,2}^2 (t)\bar{K}_{22}}{(1+r^2 \bar{g}_2)^2}
%+ \frac{1}{2} \frac{\bar{K}_{11}}{1+r^2 \bar{g}_1} 
%+ \frac{1}{2} \frac{\bar{K}_{22}}{1+r^2 \bar{g}_2} 
%\right) \nonumber\\
%
& = & \frac{S_{ss'kl}(t)}{4} \int_0^{\bar{g}_1^{-1/2}} \mathrm{d}u\,u^3\,\frac{
\mathrm{e}^{ - \frac{u^2 \eta}{1-u^2\bar{g}_1} - u^2 \left( \bar{b}_{ss'kl,1}^2 (t) \bar{g_1} + \frac{\bar{b}_{ss'kl,2}^2 (t) \bar{g_2}}{1-u^2\Delta\bar{g}} \right) }
}{\sqrt{1-u^2\Delta\bar{g}}}\,
\times\Bigg(
  \bar{K}_{11} \bar{b}_{ss'kl,1}^2 (t)
\nonumber\\&&
\hspace{-1.5cm}
+2\frac{\bar{K}_{12} \bar{b}_{ss'kl,1} (t) \bar{b}_{ss'kl,2} (t)}{1-u^2\Delta\bar{g}}
+ \frac{\bar{K}_{22} \bar{b}_{ss'kl,2}^2 (t)}{(1-u^2\Delta\bar{g})^2}
+ \frac{1}{2} \frac{\bar{K}_{11}}{(1-u^2\bar{g}_1)}
+ \frac{1}{2} \frac{\bar{K}_{22}}{(1-u^2\Delta\bar{g})(1-u^2\bar{g}_1)}
\Bigg). 
\end{eqnarray}
The last two terms diverge in the upper limit of the integral.
However, we can add and remove these terms in their upper limit to isolate the singularities and analytically integrate them:
\begin{eqnarray}\label{eq:DBOC4}
I 
& = & 
%+\frac{\pi^{\mathcal{D}/2}}{4} \mathrm{e}^{ \vert\vert\bar{\mathbf{b}}_{ss'kl} (t)\vert\vert^2 + \bar{c}_{ss'kl}^{*} (t) } 
%\left(
%  \frac{\bar{K}_{11}}{\bar{g}_1}
%+ \frac{\bar{K}_{22}}{\bar{g}_2}
%\right) \frac{
%\mathrm{e}^{ - \left( \bar{b}_{ss'kl,1}^2 (t) + \bar{b}_{ss'kl,2}^2 (t) \right) }
%}{2\sqrt{\bar{g}_2}}
%\int_0^{\bar{g}_1^{-1/2}} \mathrm{d}u\,\frac{
%\mathrm{e}^{ - \frac{u^2 \eta}{1-u^2\bar{g}_1} }
%}{1-u^2\bar{g}_1}
\frac{S_{ss'kl}(t)}{4} 
\left(
  \frac{\bar{K}_{11}}{\bar{g}_1}
+ \frac{\bar{K}_{22}}{\bar{g}_2}
\right) \frac{
\mathrm{e}^{ - \left( \bar{b}_{ss'kl,1}^2 (t) + \bar{b}_{ss'kl,2}^2 (t) \right) }
}{4\sqrt{\bar{g}_1\bar{g}_2}}
%\int_0^{\infty} \mathrm{d}x\,\frac{
%\mathrm{e}^{ - x^2 }
%}{\sqrt{\eta/\bar{g}_1-x^2}}
\mathrm{e}^{\frac{\eta}{2\bar{g}_1}}
\mathrm{K}_0 \left(\frac{\eta}{2\bar{g}_1}\right)
\nonumber\\&& 
+ \frac{S_{ss'kl}(t)}{4} \int_0^{\bar{g}_1^{-1/2}} \mathrm{d}u 
\Bigg[ u^3\,\frac{
\mathrm{e}^{ - \frac{u^2 \eta}{1-u^2\bar{g}_1} - u^2 \left( \bar{b}_{ss'kl,1}^2 (t) \bar{g_1} + \frac{\bar{b}_{ss'kl,2}^2 (t) \bar{g_2}}{1-u^2\Delta\bar{g}} \right) }
}{\sqrt{1-u^2\Delta\bar{g}}}\,
\times\Bigg(
  \bar{K}_{11} \bar{b}_{ss'kl,1}^2 (t)
\nonumber\\&&
\hspace{-0cm}
+2\frac{\bar{K}_{12} \bar{b}_{ss'kl,1} (t) \bar{b}_{ss'kl,2} (t)}{1-u^2\Delta\bar{g}}
+ \frac{\bar{K}_{22} \bar{b}_{ss'kl,2}^2 (t)}{(1-u^2\Delta\bar{g})^2}
+ \frac{1}{2} \frac{\bar{K}_{11}}{(1-u^2\bar{g}_1)}
+ \frac{1}{2} \frac{\bar{K}_{22}}{(1-u^2\Delta\bar{g})(1-u^2\bar{g}_1)}
\Bigg)
\nonumber\\&& \hspace{0cm} 
-\left(
  \frac{\bar{K}_{11}}{\bar{g}_1}
+ \frac{\bar{K}_{22}}{\bar{g}_2}
\right) \frac{
\mathrm{e}^{ - \left( \bar{b}_{ss'kl,1}^2 (t) + \bar{b}_{ss'kl,2}^2 (t) \right) }
}{2\sqrt{\bar{g}_2}}
\frac{
\mathrm{e}^{ - \frac{u^2 \eta}{1-u^2\bar{g}_1} }
}{1-u^2\bar{g}_1}
\Bigg]
,% \nonumber\\
\end{eqnarray}
where $\mathrm{K}_0$ is the Bessel function of second kind, which has the following expansion using the Euler-Mascheroni constant $\gamma$
\begin{eqnarray}\label{eq:K0}
%\mathrm{K}_0 \left(\frac{\eta}{2\bar{g}_1}\right) & = & \log(\frac{8\bar{g}_1}{\eta}) - \gamma + \frac{\eta^2}{64\bar{g}_1^2} (\log(\frac{8\bar{g}_1}{\eta}) - \gamma + 1) + \mathcal{O}(\eta^4). 
\mathrm{K}_0 \left(\frac{\eta}{2\bar{g}_1}\right) & = & \log(8\bar{g}_1) - \log(\eta) - \gamma + \frac{\eta^2}{64\bar{g}_1^2} (\log(8\bar{g}_1) - \log(\eta) - \gamma + 1) + \mathcal{O}(\eta^4). 
\end{eqnarray}
The integral in Eq.~\ref{eq:DBOC4} now converges to a finite value, and we shall focus on the first term.
It is clear that the first term on the right-hand side of Eq.~\ref{eq:K0}, $\log({8\bar{g}_1}/{\eta})$, diverges in the limit where $\eta\to0$, while the remaining terms have a finite value.
Hence, the \gls{DBOC} term is non-integrable for Gaussian functions in a linear vibronic coupling model and gives rise to a logarithmic divergence as it was found in Ref.~\citenum{Meek:2016/jcp/184109}.

%%%%%%%%%% Insert bibliography here %%%%%%%%%%%%%%

\end{document}

\title{Supplementary information for ``The moving crude adiabatic alternative to the adiabatic representation in excited state dynamics''}% 
\author{Rosa Maskri} %
\affiliation{MSME, Univ Gustave Eiffel, CNRS UMR 8208, Univ Paris Est Creteil, F-77474 Marne-la-Vall\'ee, France} %
\author{Lo{\"i}c Joubert-Doriol} %
\email{loic.joubert-doriol@univ-eiffel.fr}
\affiliation{MSME, Univ Gustave Eiffel, CNRS UMR 8208, Univ Paris Est Creteil, F-77474 Marne-la-Vall\'ee, France} %

\date{\today}

%\pacs{}

\maketitle

%\glsresetall

%%%%%%%%%%%%%%%%%%%%%%%%%%%%%%%%%%%%%%%%%%%%%%
\section{Integration scheme}
%%%%%%%%%%%%%%%%%%%%%%%%%%%%%%%%%%%%%%%%%%%%%%

Since the Gaussians parameters evolve independently of the coefficients, we integrate them separately using a second order implicit scheme similar to the Crank-Nicolson method.
The resulting set of non-linear equations is given by
\begin{eqnarray}\label{eq:CNCl}
\mathbf{Q}_{sk} (t+\delta t) - \frac{\delta t}{2}\dot{\mathbf{Q}}_{sk} (t+\delta t) = \mathbf{Q}_{sk} (t) + \frac{\delta t}{2}\dot{\mathbf{Q}}_{sk} (t), \nonumber\\
\mathbf{P}_{sk} (t+\delta t) - \frac{\delta t}{2}\dot{\mathbf{P}}_{sk} (t+\delta t) = \mathbf{P}_{sk} (t) + \frac{\delta t}{2}\dot{\mathbf{P}}_{sk} (t),
\end{eqnarray}
which is solved by using Newton's method.

The time evolution for the coefficients $C_{sk}(t)$ is obtained by using the Pad\'e approximant to the solution of the Schr\"odinger equation
\begin{eqnarray}\label{eq:PadeSE}
\left( 1 + \imag \hat H \delta t/2 \right) \ket{\Psi(t+\delta t)} & = & \left( 1 - \imag \hat H \delta t/2 \right) \ket{\Psi(t)}.
\end{eqnarray}
Since we already know the basis from the integration of $\mathbf{Q}_{sk}(t)$ and $\mathbf{P}_{sk}(t)$, we can project Eq.~\ref{eq:PadeSE} on the basis at time $t$ and $t+\delta t$.
For the sake of generality, we introduce a unique symbol for all types of electron-nuclear basis $\ket{\phi_{sk} (t)}$ such that $\ket{\Psi(t)}=\sum_{sk}C_{sk}(t)\ket{\phi_{sk}(t)}$ (for example, in the \gls{MCA} representation $\ket{\phi_{sk} (t)}=\ket{\varphi_{sk} (\mathbf{Q}_{sk}(t))} g(\mathbf{Q}_{sk}(t),\mathbf{P}_{sk}(t))$).
We then obtain a set of two linear equations forming an overcomplete system:
\begin{widetext}
\begin{eqnarray}\label{eq:ProjPadeSE}
\left[ \mathbf{S}(t,t+\delta t) + \frac{\imag\delta t}{2} \mathbf{H}(t,t+\delta t) \right] \mathbf{C}(t+\delta t) 
& = & \left[ \mathbf{S}(t,t) - \frac{\imag\delta t}{2} \mathbf{H}(t,t)  \right] \mathbf{C}(t), \\
\left[ \mathbf{S}(t+\delta t,t+\delta t) + \frac{\imag\delta t}{2} \mathbf{H}(t+\delta t,t+\delta t) \right] \mathbf{C}(t+\delta t) & = & \left[ \mathbf{S}(t+\delta t,t) - \frac{\imag\delta t}{2} \mathbf{H}(t+\delta t,t) \right] \mathbf{C}(t),
\end{eqnarray}
\end{widetext}
where
\begin{eqnarray}\label{eq:Hmat}
S_{ss',kl}(t,t') & = & \braket{\phi_{sk}(t)}{\phi_{s'l}(t')}, \\
H_{ss',kl}(t,t') & = & \braOket{\phi_{sk}(t)}{\hat H}{\phi_{s'l}(t')}.
\end{eqnarray}
We then solve for $\mathbf{C}(t+\delta t)$ by applying the least-square procedure.
Since this procedure does not conserve the norm of the wave-function, the quantity $\mathbf{C}(t+\delta t)$ is then rescaled so that $\mathbf{C}(t+\delta t)^\dagger\mathbf{S}(t+\delta t)\mathbf{C}(t+\delta t)=1$.
This rescaling factor do not impact further analysis of the results since all expectation values do not depend on this factor, but it allows for more stability of the propagation.

The integration time step is $5\cdot10^{-2}$ fs for both the Gaussian parameters and the coefficients $C_{sk}(t)$.

All integrals needed in Eq.~\ref{eq:Hmat} are calculated exactly using Gaussian integration and the Gamma function definition $\Gamma(y)$ to integrate rational functions of the type $f(\mathbf{X})^{-y}$ where $f(\mathbf{X})$ is any positive function of the coordinates $\mathbf{X}$.
\begin{eqnarray}\label{eq:Gamma}
\Gamma(y) f(\mathbf{X})^{-y} & = & 2 \int_0^\infty \mathrm{d}r\, r^{2y-1} \mathrm{e}^{-f(\mathbf{X})r^2}.
\end{eqnarray}
The \gls{MCA} simulations generate results that are then rotated to the adiabatic representation using a similar integration technique.

%%%%%%%%%%%%%%%%%%%%%%%%%%%%%%%%%%%%%%%%%%%%%%
\section{Spawning procedure}
%%%%%%%%%%%%%%%%%%%%%%%%%%%%%%%%%%%%%%%%%%%%%%

The spawning procedure~\cite{Curchod:2020/Book} is activated when a Gaussian's center $\mathbf{Q}_{sk}(t)$ enters a region of strong non-adiabatic coupling, i.e. when $||\boldsymbol{\tau}_{12}(\mathbf{Q}_{sk}(t))||>2\cdot10^1$.
Once this criterion is fulfilled, the center of the original, parent, Gaussian is propagated all the way till it exits the region.
A spawn of the parent Gaussian, the child Gaussian, is then created on the other surface.
When it is possible, the momentum of the created Gaussian is rescaled along the non-adiabatic coupling vector so that the energy of the parent Gaussian is the same as the child one.
If this is not possible, the momentum of the child Gaussian is optimized to minimize the energy difference between the parent and the child Gaussians.
Once the new child basis is created, it is propagated backward in time until the time at which the parent Gaussian entered the spawning region. This time is then also the time from where the child Gaussian can be populated.

%%%%%%%%%%%%%%%%%%%%%%%%%%%%%%%%%%%%%%%%%%%%%%
\section{Exact calculation}
%%%%%%%%%%%%%%%%%%%%%%%%%%%%%%%%%%%%%%%%%%%%%%

Gaussian-based calculations are compared to exact calculations using a finite basis representation of the Hamiltonian $\mathbf{H}^\mathrm{dia}$ of the main text.
The nuclear components are expanded in a direct product basis of harmonic oscillator eigenfunctions centered at the origin of the coordinate system with $55$ basis functions in each direction.
We then employ the spectral decomposition with an energy cut off of the basis at $30\times\langle\Psi\vert\hat H \Psi\rangle$ for the time-dependent integration.
For comparison purpose, the exact time-dependent wavefunction is then rotated to adiabatic representation after transforming to a discrete variable representation.

All simulations are done with the Octave package~\cite{octave}.